\documentclass[conference]{IEEEtran}
\IEEEoverridecommandlockouts

\usepackage{cite}
\usepackage[numbers,sort&compress]{natbib}
\usepackage{amsmath,amssymb,amsfonts}
\usepackage{graphicx}
\usepackage{subfigure}
\usepackage{textcomp}
\usepackage{xcolor}
\usepackage{bm}

\usepackage[noend]{algpseudocode}
\usepackage{algorithmicx,algorithm}

\usepackage{amsthm}

\usepackage{stfloats}
\usepackage{amsmath}

\def\BibTeX{{\rm B\kern-.05em{\sc i\kern-.025em b}\kern-.08em
    T\kern-.1667em\lower.7ex\hbox{E}\kern-.125emX}}

\ifCLASSINFOpdf
\else
\fi

\begin{document}
\linespread {0.96}  
\addtolength{\parskip}{0.98pt} 

\setlength{\columnsep}{10.6pt}

\makeatletter
\renewcommand\normalsize{%
\@setfontsize\normalsize\@xpt\@xiipt
\abovedisplayskip 4\p@ \@plus2\p@ \@minus5\p@
\abovedisplayshortskip \z@ \@plus3\p@
\belowdisplayshortskip 6\p@ \@plus3\p@ \@minus3\p@
\belowdisplayskip \abovedisplayskip
\let\@listi\@listI}
\makeatother

\title{Doubly-Dynamic ISAC Precoding for Vehicular Networks: A Constrained Deep Reinforcement Learning (CDRL) Approach\\

}


\author{\IEEEauthorblockN{ Zonghui Yang$^*$, Shijian Gao$^\dagger$, Xiang Cheng$^*$}
\IEEEauthorblockA{$^*$State Key Laboratory of Advanced Optical Communication
Systems and Networks, \\
School of Electronics, Peking University, Beijing, China.\\
$^\dagger$Internet of Things Thrust, The Hong Kong University of Science and Technology (Guangzhou), Guangzhou, China.\\}
Email: yzh22@stu.pku.edu.cn, shijiangao@hkust-gz.edu.cn, xiangcheng@pku.edu.cn
}


\maketitle
\vspace{-0.6cm}
\begin{abstract}


Integrated sensing and communication (ISAC) technology is essential for supporting vehicular networks. However, the communication channel in this scenario exhibits time variations, and the potential targets may move rapidly, resulting in double dynamics. This nature poses a challenge for real-time precoder design. While optimization-based solutions are widely researched, they are complex and heavily rely on perfect channel-related information, which is impractical in double dynamics. To address this challenge, we propose using constrained deep reinforcement learning to facilitate dynamic updates to the ISAC precoder. Additionally, the primal dual-deep deterministic policy gradient and Wolpertinger architecture are tailored to efficiently train the algorithm under complex constraints and varying numbers of users. The proposed scheme not only adapts to the dynamics based on observations but also leverages environmental information to enhance performance and reduce complexity. Its superiority over existing candidates has been validated through experiments.

\end{abstract}

\IEEEpeerreviewmaketitle

\vspace{-0.55cm}
\section{Introduction}

The Internet of Vehicles (IoV) is essential for supporting an efficient and reliable intelligent transportation system, with high-speed communications and high-precision sensing serving as cornerstones \cite{overview_JRC_for_AV,cheng2022integrated}. In comparison to separate communication and sensing systems, integrated sensing and communication (ISAC) offers the potential for mutual enhancement, making it a recent research focus in vehicular networks \cite{Joint_opt_ISAC, radar_integrated_V2V_Fan, ISAC_CRB_opt,overview_Som, MMFF_Som}.
However, the dynamic nature of the road environment introduces time-varying characteristics in the wireless channel due to user and scatterer movements. Simultaneously, the targets themselves move rapidly and unpredictably, presenting a doubly-dynamic challenge for ISAC.

While previous studies by \cite{dynamic_com_1, dynamic_com_2} and \cite{ dynamic_sensing_2} have considered ISAC under individual communication or sensing dynamics, implementing ISAC under double dynamics remains unexplored. Compared to single dynamics, the presence of double dynamics introduces multiple challenges to the design of ISAC precoding. Firstly, existing optimization-based methods, e.g., \cite{Joint_opt_ISAC,ISAC_CRB_opt}, exhibit high computational complexity, resulting in unaffordable latency. Secondly, Doppler shifts in time-varying channels can induce severe inter-carrier interference (ICI) in wideband systems, leading to degraded spectrum efficiency (SE) and sensing accuracy.  Moreover, applying optimization-based methods typically requires accurate prior information, including instantaneous channel state information (CSI) and real-time target's position prior, which can be prohibitively challenging to acquire in double dynamics.

In an effort to mitigate the complexity issue and lower the dependence on real-time prior information, methods such as message passing (MP) and particle filter (PF) with modified operating protocols have been proposed in \cite{dynamic_MP, dynamic_PF}. However, these solutions still struggle to address the doubly-dynamics challenge, since the mutual interference between communication and sensing can degrade the performance of these algorithms. Furthermore, as the number of users increases, the complexity of these methods increases significantly, diminishing the timeliness of ISAC systems.

Reinforcement Learning (RL), known for its adaptability to dynamics without prior assumptions, has garnered attention in applications to ISAC systems \cite{RL_radar_ISAC, RL_time_divide}. However, 
applying RL to the considered doubly-dynamic problem still faces two main difficulties: Firstly, existing RL algorithms in \cite{SAC, DDPG} struggle to handle complex constraints in this problem, resulting in performance degradation. Secondly, the parameter tuning in precoding design entails a large action space, the size of which may vary with the number of users, posing difficulty in action selection and policy updates.

To overcome the aforementioned barriers, we devise an efficient approach via constrained deep-RL (CDRL) to facilitate real-time wideband precoding design for instantaneous target tracking and multi-user communications, even in the absence of perfect prior knowledge.
This approach incorporates various state observations, such as historical position estimates and initial CSI, which serve as available inputs to adapt to the double dynamics.
To address the intricate constrained problem, we formulate a constrained Markov decision process (CMDP) and employ the primal-dual deep deterministic policy gradient (PD-DDPG) algorithm for effective training. Additionally, to further accommodate the changing number of users in practical applications, we have customized a Wolpertinger-based scheme for flexible action selection. Numerical experiments have substantiated the superiority of the proposed scheme in terms of both communications and target tracking.


\newcommand{\RNum}[1]{\uppercase\expandafter{\romannumeral #1\relax}}
\vspace{-0.2cm}
\section{System Model}
\vspace{-0.2cm}
We consider an ISAC system with wideband massive MIMO at the base station (BS) serving $U$ downlink single-antenna users, while tracking one low-altitude moving target in the environment, as illustrated in Fig.~1. The BS is equipped with a uniform linear array (ULA) comprising $N_{\text{t}}$ elements for transmission and $N_{\text{r}}$ elements for echo reception. The ISAC signal is transmitted via orthogonal frequency division multiplexing (OFDM) with $M$ subcarriers.
Each frame is divided into $T$ subframes, each having $L$ OFDM symbols. The positions of the targets and users are assumed to be invariant per subframe and the velocities remain constant within one frame.

\vspace{-0.2cm}
\subsection{Transmitted Waveform}
Hybrid precoding is adopted at BS, including a frequency-independent analog precoder and $M$ frequency-dependent digital precoders. Here we simply assume the digital precoders to be $\bm I_{\text{U}}$, leaving the extension of their optimization as future work. Thus for the $l$-th symbol in the $t$-th subframe, the transmitted signal at the $m$-th subcarrier can be expressed as
\vspace{-0.2cm}
\begin{equation}
    \bm x_{\text{m}}^{(t,l)}=\sqrt{\frac{P_{\text{t}}}{N_{\text{t}}UM}}\bm F^{(t)}\bm s_{\text{m}}^{(t,l)},
    \vspace{-0.3cm}
\end{equation}
where $P_{\text{t}}$ denotes the transmitting power, $\bm s_{\text{m}}^{(t,l)}=\left[ s_{\text{m},1}^{(t,l)},\cdots, s_{\text{m},\text{U}}^{(t,l)}\right ]^{\text{T}}$ contains the communication symbols for $U$ users at the $m$-th subcarrier with $\mathbb{E} \left [\bm s_{\text{m}}\bm s_{\text{m}}^{\text{H}}\right ]=\bm I_{\text{U}}$, and $\bm F^{(t)} \in \mathbb{R}^{N_{\text{t}}\times \text{U}}$ is the precoder at each subcarrier, updated at a subframe-level timescale. Since the analog precoder is composed of phase shifters between radio frequency (RF) chains and the antenna array, each element of $\bm F$ has constant modulus, i.e., $\vert [\bm F]_{i,j} \vert^{2}=1$. 

\vspace{-0.2cm}
\subsection{Communication Model}
The doubly-selective channel is adopted similar to \cite{DSDS_TWC}. At the $d$-th tap $(0\leqslant d < N_{\text{d}})$, the channel response for user-$u$ during symbol-$l$ is 
\vspace{-0.2cm}
\begin{equation}
\widetilde{\bm h}_{\text{u},\text{d}}(l)=\sqrt{\frac{N_{\text{t}}}{P_{\text{u}}}}\sum_{p=1}^{P_{\text{u}}}\beta_{\text{u},\text{p}}g_{\text{rp}}(dT_{\text{s}}-\tau_{\text{u},\text{p}})e^{j\omega_{\text{u},\text{p}}l} \bm a_{\text{t}}(\theta_{\text{u},\text{p}}),
\vspace{-0.2cm}
\end{equation}
where $P_{\text{u}}$ denotes the number of resolvable paths, $T_{\text{s}}$ is the sampling period, $\beta_{\text{u},\text{p}}\sim\mathcal{CN}(0,\sigma_{\beta}^{2})$ is the path's complex gain and $\theta_{\text{u},\text{p}}$ is the angle of departure (AoD), with the $i$-th element $(i=1,\cdots,N_{\text{t}})$ of $\bm a_{\text{t}}(\theta)$ being $\frac{1}{\sqrt{N_{\text{t}}}}e^{-j(i-1)\pi \sin{\theta}}$.
$g_{\text{rp}(\cdot)}$ is the pulse shaping filter; $\tau_{\text{u},\text{p}}\sim \mathcal{U}(0,(N_{\text{d}}-1)T_{\text{s}})$ is the propagation delay. Denote the carrier frequency as $f_{\text{c}}$, the velocity as $c_{\text{v}}$, and the relative velocity as $\bm v_{\text{u},\text{p}}$, then the normalized Doppler shift is $\omega_{\text{u},\text{p}}=2\pi f_{\text{c}}\vert \bm v_{\text{u},\text{p}}\vert T_{\text{s}}\sin{(\theta_{\text{u},\text{p}})}/c_{\text{v}}$.

The multi-user time-domain channel is $\widetilde{\bm H}_{\text{d}}(l)=\left[\widetilde{\bm h}_{1,\text{d}}(l), \cdots, \widetilde{\bm h}_{\text{U},\text{d}}(l) \right]$. The received frequency-domain signal $\bm y_{\text{c}}\in \mathbb{C}^{\text{UM}\times 1}$ across all subcarriers is
\vspace{-0.2cm}
\begin{align}
&\bm y_{\text{c}}=\left[  \bm y_{1}^{\text{T}}, \bm y_{2}^{\text{T}}, \cdots, \bm y_{\text{M}}^{\text{T}} \right]^{\text{T}} \\ \nonumber
&=
\sqrt{\frac{P_{\text{t}}}{N_{\text{t}}UM}}\!\underbrace{
\left[
\begin{matrix}
\bm H_{\text{1}}[1] \! &\! \cdots\! & \!\bm H_{\text{1}}[M]\\
\vdots \!& \!\ddots \!&\! \vdots \\
\bm H_{\text{M}}[1] \!&\! \cdots \!& \!\bm H_{\text{M}}[M]
\end{matrix}
\right]
}_{\overline{\bm H}}
\!\underbrace{
\left[ 
\begin{matrix}
\bm F_{\text{1}} \!&\! \! &\!    \\
 &\!   \ddots \!&\!  \\
 &\! \!&\! \bm F_{\text{M}} \\
\end{matrix}
\right]
}_{\overline{\bm F}}
\!\left[
\begin{matrix}
\bm s_{\text{1}}\\ \vdots \\ \bm s_{\text{M}}
\end{matrix}
\right]\!\!+\!\!\bm n_{\text{c}}.
\vspace{-0.4cm}
\end{align}
In Eq.~(3), $\bm H_{\text{m}}[k]$ stands for the frequency-domain channel response from the $k$-th to the $m$-th subcarrier, through the Fourier transform on $\widetilde{\bm H}_{\text{d}}(l)$ \cite{fan2021wideband}, and $\bm n_{\text{c}}\!\sim \! \mathcal{CN}(\bm 0,\sigma_{\text{c}}^{2}\bm I_{\text{UM}})$ is the white Gaussian noise.
In the studied context, $\beta_{\text{u},\text{p}}$, $\tau_{\text{u},\text{p}}$ and $\theta_{\text{u},\text{p}}$ are varying across subframes as the users and related scatterers move, based on their current velocities.

Note that at the $m$-th subcarrier, the $u$-th user receives not only the customized communication signals but also the inter-user interference (IUI) and ICI. Hence, its signal-to-interference-plus-noise ratio (SINR) is
\vspace{-0.2cm}
\begin{equation}
    \gamma_{\text{u},\text{m}}^{(t,l)}\!\!=\!\frac{|\bm h_{\text{m},\text{m},\text{u}}^{(t,l)}\bm f_{\text{u}}^{(t)}|^{2}}{\!\sum_{k\!\neq \!m}\!\sum_{u\!=\!1}^{U}\!|\bm h_{\text{m},\text{k},\text{u}}^{(t,l)}\bm f_{\text{u}}^{(t)}\!|^{2}\!+\!\sum_{i \!\neq \! u}\!|\bm h_{\text{m},\text{m},\text{u}}^{(t,l)}\bm f_{\text{i}}^{(t)} \!|^{2} \!+\!\!\frac{N_{\text{t}}UM\!\sigma_{\text{c}}^{2}}{P_{\text{t}}\!}},
\end{equation}
where $\bm f_{\text{u}}=\bm F[:, u]$ and $\bm h_{\text{m},\text{k},\text{u}}=\bm H_{\text{m}}[k][u,:]$. Therefore, the SE can be calculated as
\vspace{-0.2cm}
\begin{equation}
 \text{SE}^{(t)}=\frac{\sum_{l=1}^{L}\sum_{m=1}^{M}\sum_{u=1}^{U}\log_{2}(1+\gamma_{\text{u},\text{m}}^{(t,l)})}{LT_{\text{s}}M \Delta f}.
\end{equation}

\begin{figure}[t]
  \vspace{-0.4cm}
  \setlength{\abovecaptionskip}{-0.2cm} 
  \setlength{\belowcaptionskip}{-1cm} 
  \centering
  \includegraphics[width=0.88 \linewidth]{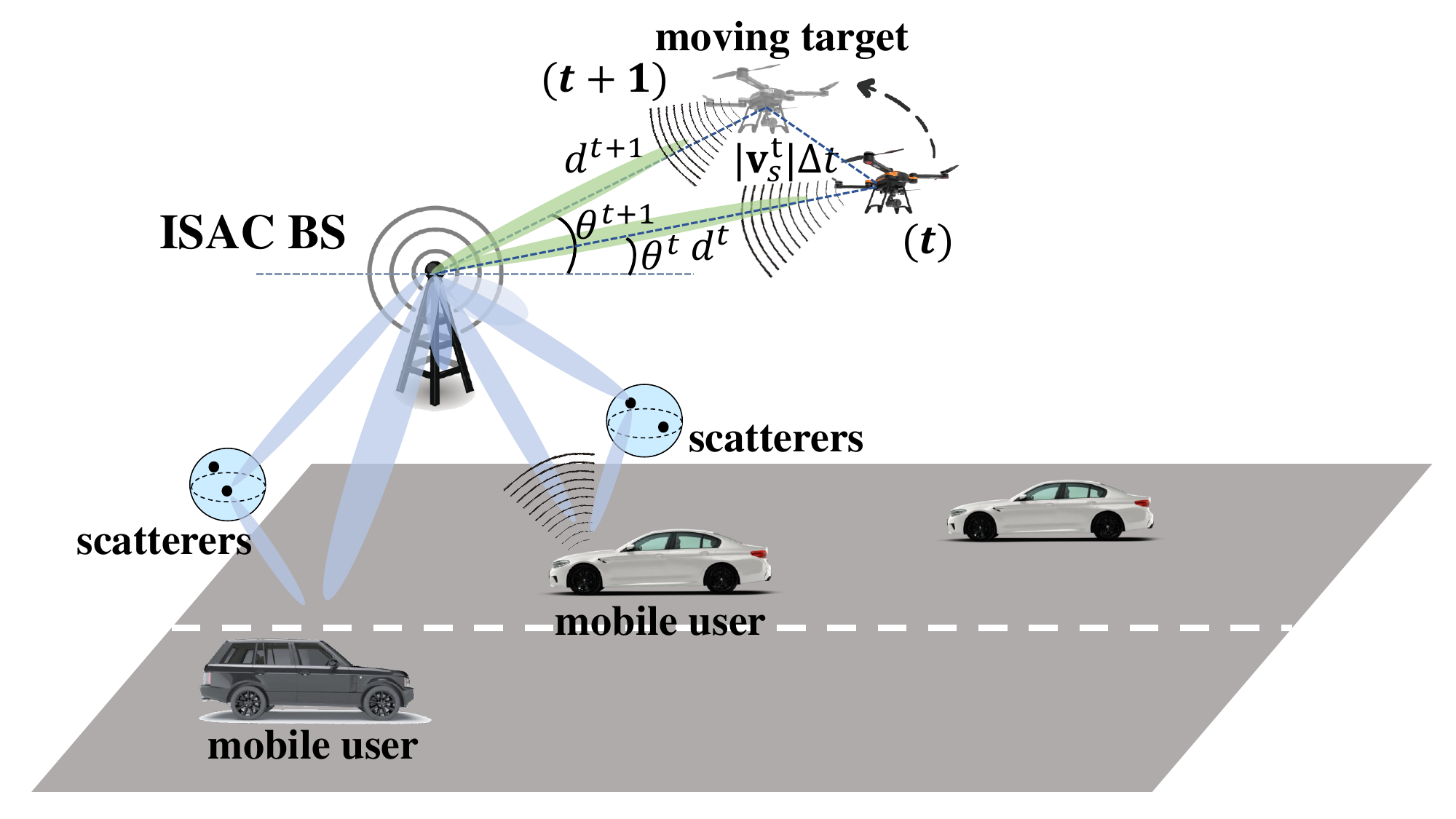}
  \vspace{0cm}
  \caption{An illustration of the ISAC system in doubly-dynamic scenarios.}
  \label{fig:system}
  \vspace{-0.65cm}
\end{figure}

\vspace{-0.2cm}
\subsection{Sensing Model}

Denote $\left[ \theta_{\text{s}}^{t},d_{\text{s}}^{t} \right ]$ is the target's position in the polar domain in the $t$-th subframe.
Define $\bm G^{(t)}_{\text{m}}=\alpha^{(t)}_{\text{m}} \bm a_{r}(\theta_{\text{s}}^{t})\bm a_{\text{t}}^{\text{H}}(\theta_{\text{s}}^{t})$ to be the target response matrix at the $m$-th subcarrier, with $\alpha_{\text{m}}^{(t)}$ being the complex reflection coefficient that incorporates the round-trip path loss, the radar cross-section (RCS) and the Doppler of the target. Accordingly, the received echo becomes
\vspace{-0.2cm}
\begin{equation}
    \bm y_{\text{r},\text{m}}^{(t,l)}=\bm G_{\text{m}}^{(t)} \bm x_{\text{m}}^{(t,l)}+\bm z_{\text{m}}^{(t,l)},
    \vspace{-0.1cm}
\end{equation}
with $\bm z_{\text{m}}^{(t,l)}\sim\mathcal{CN}(\bm 0, \sigma_{\text{z}}^{2}\bm I_{N_r})$ being the ISAC receiver noise.

\newcounter{TempEqCnt} 
\setcounter{TempEqCnt}{\value{equation}} 
\setcounter{equation}{6} 
\vspace{-0.1cm}
\begin{figure*}[ht]
\begin{equation}
\label{equ:CRB}
     \mathcal{J}_{\text{m}}(\theta_{\text{s}}^{t})=\frac{2\| \alpha_{\text{m}} \|^{2}(\text{Tr}(\dot{\bm A} ^{\text{H}}(\theta_{\text{s}}^{t})\dot{\bm A}(\theta_{\text{s}}^{t})\bm R_{\text{x},\text{m}}^{(t)})\text{Tr}(\bm A^{\text{H}}(\theta_{\text{s}}^{t})\bm A(\theta_{\text{s}}^{t})\bm R_{\text{x},\text{m}}^{(t)})-\|\text{Tr}(\dot{\bm A}^{\text{H}}(\theta_{\text{s}}^{t}) \bm A(\theta_{\text{s}}^{t}) \bm R_{\text{x},\text{m}}^{(t)} )\|^{2})} {\sigma_{\text{z}}^{2}\text{Tr}(\bm A^{\text{H}}(\theta_{\text{s}}^{t})\bm A(\theta_{\text{s}}^{t})\bm R_{\text{x},\text{m}}^{(t)})}.
\end{equation}
\hrulefill
\vspace{-0.6cm}
\end{figure*}

The sensing accuracy is characterized by Fisher information in Eq.~(\ref{equ:CRB}) at the top of the next page, with $\bm R_{\text{x},\text{m}}^{(t)}=\frac{1}{L}\mathbb{E}\left[ \sum_{l=1}^{L} \bm x_{\text{m}}^{(t,l)}\bm x_{\text{m}}^{(t,l),\text{H}} \right ] =\frac{P_{\text{t}}}{N_{\text{t}}UM}\bm F^{(t)}\bm F^{(t),\text{H}}$ being the covariance matrix of $\bm x_{\text{m}}$, and $\bm A(\theta_{\text{s}})=\bm a_{\text{r}}(\theta_{\text{s}})\bm a_{\text{t}}^{\text{H}}(\theta_{\text{s}})$ with $ \dot{\bm A}(\theta_{\text{s}})=\frac{\partial \bm A(\theta_{\text{s}})}{\partial \theta_{\text{s}}}$. The Fisher information for $\theta_{\text{s}}^{(t)}$ in wideband systems is the summation of all the Fisher information across subcarriers. Consequently, the Cramér-Rao lower bound (CRLB) of the target's angle is the expectation of the inverse of Fisher information given $\bm y_{r}$.
\vspace{-0.2cm}
\begin{equation}
    \text{CRLB}(\theta_{\text{s}}^{t})=\mathbb{E}\left [\frac{1}{\mathcal{J}(\theta_{\text{s}}^{t})}\right ]=\mathbb{E}\left [\frac{1}{\sum_{m=1}^{M}\mathcal{J}_{\text{m}}(\theta_{\text{s}}^{t})}\right ].
\end{equation}
\vspace{-0.3cm}

Accordingly, the averaged CRLB over time, i.e., $\overline{\text{CRLB}}=\frac{1}{T}\sum_{t=1}^{T}\text{CRLB}(\theta_{\text{s}}^{t})$, is established as the sensing performance indicator to optimize the ISAC precoder.

\vspace{-0.2cm}

\section{Optimization-based ISAC precoding \\ with perfect prior}
From the communication-centric perspective, the objective of the doubly-dynamic ISAC precoding is to minimize the averaged CRLB for angle estimation, while also ensuring SE throughout the frames, and taking into account the constant-modulus constraint of the precoder. Therefore, the corresponding problem can be formulated as
\vspace{-0.25cm}
\setlength{\jot}{-0.01em} 
\begin{align}
    \underset{ \{\bm F^{(t)}\}_{t=1}^{T} }{\min}&~ \overline{\text{CRLB}}=\frac{1}{T}\sum_{t=1}^{T}\text{CRLB}(\theta_{\text{s}}^{t}) \label{Problem0}\\
    \text{s.t.} ~~&~ \gamma_{\text{u},\text{m}}^{(t,l)}\geqslant \tau, \forall ~u \in [1,U], \forall ~m\in [1,M], \tag{\ref{Problem0}{a}} \label{Problem0_a}\\
    &~ \vert [\bm F^{(t)}]_{i,j} \vert=1,\tag{\ref{Problem0}{b}}\label{Problem0_b}
    \vspace{-0.5cm}
\end{align}
given the perfect $\overline{\bm H}$ and perfect $\bm G_{\text{m}}$ in each subframe.
Solving this optimization problem directly can be tricky due to the complex objective function and constant-modulus constraints. Furthermore, the ICI induced by Doppler imposes a more complex constraint in (\ref{Problem0_a}) than in \cite{ISAC_CRB_opt}.
To address these challenges, we approach the sequential optimization on a subframe-by-subframe basis as
\vspace{-0.25cm}
\begin{align}
    \underset{ {\bm F}}{\min}&~ \text{CRLB}(\theta_{\text{s}}) \label{Problem1}\\
    \text{s.t.} &~ \gamma_{\text{u},\text{m}}\geqslant \tau, \forall ~u \in [1,U], \forall ~m\in [1,M], \tag{\ref{Problem1}{a}} \label{Problem1_a}\\
    &~ \vert [\bm F]_{i,j} \vert=1. \tag{\ref{Problem1}{b}}\label{Problem1_b}
    \vspace{-0.6cm}
\end{align}
Subsequently, we employ the semi-definite relaxation (SDR) technique to iteratively derive a sub-optimal solution at $t$. Denote $\bm W_{\text{u}}=\bm f_{\text{u}}\bm f_{\text{u}}^{\text{H}}$ and $\bm Q_{\text{m},\text{k},\text{u}}=\bm h_{\text{m},\text{k},\text{u}}^{\text{H}}\bm h_{\text{m},\text{k},\text{u}}$. Then we reformulate the objective and constraints as
\vspace{-0.25cm}
\begin{align}
     \underset{\bm W, \{\bm W_{\text{u}}\}_{u=1}^{U} }{\max} & ~~~\zeta \label{Problem2}\\
    \text{s.t.}~~~~ &~ \left [  \begin{matrix}  \text{Tr}(\Dot{\bm A}^{\text{H}}\Dot{\bm A}\bm W)-\zeta  &  \text{Tr}(\Dot{\bm A}^{\text{H}}\bm A\bm W) \\  \text{Tr}(\bm A^{\text{H}}\Dot{\bm A}\bm W) &  \text{Tr}(\bm A^{\text{H}}\bm A\bm W)  \end{matrix}   \right] \succeq \bm 0, \tag{\ref{Problem2}{a}} \label{Problem2_a}\\
    &~ \text{Tr}(\bm Q_{\text{m},\text{m},\text{u}}\bm W_{\text{u}})-\tau \sum_{i=1,i\neq u}^{\text{U}}\!\text{Tr}(\bm Q_{\text{m},\text{m},\text{u}}\bm W_{\text{i}})  \nonumber \\
    &-\tau \sum_{k=1, k\neq m}\bm Q_{\text{m},\text{k},\text{u}}\bm W \leqslant \tau \sigma_{\text{c}}^{2}, \tag{\ref{Problem2}{b}} \label{Problem2_b}\\
    &~ \text{Tr}(\bm E_{\text{i}}^{\text{H}}\bm W_{\text{u}})\!=\!1, \forall u \in [1,U], \forall i\in [1,N_{\text{t}}],  \tag{\ref{Problem2}{c}} \label{Problem2_c}\\ \nonumber
    &~ \bm W=\sum_{u=1}^{\text{U}}\bm W_{\text{u}}, \tag{\ref{Problem2}{d}} \label{Problem2_d}\\ \nonumber
    &~ \bm W_{\text{u}}\succeq \bm 0, ~~\text{rank}(\bm W_{\text{u}})=1, \tag{\ref{Problem2}{e}} \label{Problem2_e} \nonumber
\vspace{-0.9cm}
\end{align}
where $\zeta$ represents for the lower bound of Fisher information, $\bm E_{\text{i}}=\bm e_{\text{i}}\bm e_{\text{i}}^{\text{T}}$ denotes the element-wise selection matrix, with $\bm e_{\text{i}}$ having 1 at the $i$-th element and 0 elsewhere. By dropping the rank-1 constraint in (\ref{Problem2_e}), the problem is relaxed to a convex one and can be solved by off-the-shelf tools.
Then the post-processing can be applied according to \cite{Joint_opt_ISAC} to satisfy the rank-1 constraint, and the precoding matrix $\bm F$ is obtained per subframe, thus improving the averaged performance in frames.

\textbf{Discussion on the applicability}: This optimization-based method solves the problem at the cost of high computational complexity, scaling up to $\mathcal{O}(N_{\text{t}}^{3.5}\log(1/\epsilon))$, given a solution accuracy $\epsilon$. This results in an unacceptable delay, especially when the antenna array becomes large, as is the case in the studied scenario.
Furthermore, in the context of the double dynamics, obtaining prior information $\bm H^{(t)}$ and $\theta_{\text{s}}^{(t)}$ for optimizing $\bm F^{(t)}$ is not readily available. Consequently, the optimization-based method may no longer be applicable.

\vspace{-0.2cm}
\section{Constrained DRL-based Precoding for ISAC without perfect prior}
Since perfect prior can hardly be acquired in strong dynamic scenarios, henceforth, we introduce a CDRL-based approach for ISAC precoding.
The problem will be formulated as a CMDP with appropriate learning strategies for efficient training.


\vspace{-0.2cm}
\subsection{Constrained Markov Decision Process Formulation}
Due to the evolving characteristics of wireless channels and positions, the doubly-dynamic scenario is initially modeled as a CMDP.
Once the environment state is observed in subframe $t$, BS takes an action according to its strategy. Subsequently, BS receives a reward as well as a cost, and the environment evolves to the next state with a certain transition probability. This process continues, with the new state being observed at each subframe, and new actions are taken accordingly. The essential elements of this CMDP are described below in detail.

\subsubsection{State Space}
The observed state at BS can be expressed as $\bm s^{(t)}\!\!=\!\!\{\bm S_{\text{H}}^{(t)},\! \bm S_{\text{P}}^{(t)}\}$, including both channel-related and position-related environmental information. 
To be specific, $\bm S_{\text{H}}^{(t)}\!=\![\vert \text{vec}(\widehat{\bm H}_{0}^{(0)}\! \bm D_{\text{t}})\vert,\! \vert\text{vec}(\widehat{\bm H}_{1}^{(0)}\! \bm D_{\text{t}})\vert, \!\cdots,\!\vert\text{vec}(\widehat{\bm H}_{N_{\text{d}}\!-\!1}^{(0)}\! \bm D_{\text{t}})\vert]$, where $\widehat{\bm H}_{\text{d}}^{(0)}$ is the estimate of the $d$-th tap channel $\widetilde{\bm H}_{\text{d}}^{(0)}(0)$ which can be obtained at the beginning of each frame. $\text{vec}(\cdot)$ and $\vert(\cdot)\vert$ denote the vectorization and the modulo operation, and $\bm D_{\text{t}}\!\in \!\mathbb{C}^{N_{\text{t}}\times G_{\text{t}}}$ is the angular dictionary to transform initial CSI into a sparser representation in beamspace and delay domain \cite{DSDS_TWC}.
We represent $\bm S_{\text{P}}^{(t)}$ using a three-dimensional time-angle-range spectrum, i.e., $\bm S_{\text{P}}^{(t)}=\left[\bm P^{(t-3)}, \bm P^{(t-2)}, \bm P^{(t-1)} \right]$, where $\bm P^{(t)}\!\in \!\mathbb{R}^{N_{\text{x}}\times N_{\text{y}}}$ denotes the angle-range spectrum estimated at the $t$-th subframe, divided into $N_{\text{x}}\times N_{\text{y}}$ grids, with a sensing range $[\theta_{\text{min}}, \theta_{\text{max}}) \times [d_{\text{min}}, d_{\text{max}})$. Specifically,
\vspace{-0.2cm}
\begin{equation}
  \bm P^{(t)}(n_{\text{x}},\! n_{\text{y}})\! = \!\!
  \begin{cases}
    1, \!\!&\text{if $\{[\theta_{n_{\text{x}}},\!\theta_{n_{\text{x}}\!+\!1})\!\times \! [d_{n_{\text{y}}},\! d_{n_{\text{y}}\!+\!1})\}\! \cap \mathcal{U}^{(t)} \!\!\neq \!\varnothing$},\\
	2,\!\! &\text{if $\{[\theta_{n_{\text{x}}},\!\theta_{n_{\text{x}}\!+\!1})\!\times \! [d_{n_{\text{y}}},\! d_{n_{\text{y}}\!+\!1})\}\! \cap \mathcal{T}^{(t)} \!\!\neq \!\varnothing$},\\
    0,\!\! &\text{Otherwise},
  \end{cases}
  \vspace{-0.2cm}
  \nonumber
\end{equation}
where $\theta_{n_{\text{x}}}\!=\!\theta_{\text{min}}\!+\!\frac{\theta_{\text{max}}\!-\!\theta_{\text{min}}}{N_{\text{x}}}(n_{\text{x}}-1)$ and $d_{n_{\text{y}}}\!=\!d_{\text{min}}\!+\frac{d_{\text{max}}-d_{\text{min}}}{N_{\text{y}}}(n_{\text{y}}-1)$. $\mathcal{U}^{(t)}$ and $\mathcal{T}^{(t)}$ are the sets of angle-range pairs of users and targets at $t$, which can be identified through BS-end echo processing and periodly-uploaded positions of users \cite{location_aware_overview}. It is worth noting that this representation of position state can adapt to varying numbers of users and targets.


\subsubsection{Action space}
Since the environment generally dose not change drastically across adjacent subframes, we propose to update $\bm F$ progressively. In subframe $t$, only one column of $\bm F^{(t-1)}$ is substituted by one codeword from the DFT codebook $\mathcal{B}$ with $\vert \mathcal{B}\vert=N_{\text{B}}$. 
The action comprises the substituted column's index in $\bm F^{(t-1)}$ and the codeword's index in $\mathcal{B}$. These are represented by $\bm a^{(t)}\in [0,1]^{N_{\text{A}}}$, where $N_{\text{A}}=\lceil \log_{2}(U^{\text{max}})\rceil+\lceil \log_{2}(N_{\text{B}})\rceil$, with $U^{\text{max}}$ being the maximal number of users.
However, the network output $\bm a^{(t)}$ may not bear the binary form to represent the integer indices. Therefore the post-processing of $\bm a^{(t)}$ is necessary, as detailed in the subsequent subsection.

\subsubsection{Reward and cost functions}
To guide BS to minimize the averaged CRLB while satisfying the SE constraint, we design the immediate reward and cost function respectively as
\vspace{-0.2cm}
\begin{equation}
r^{(t)}=-\text{CRLB}(\theta_{\text{s}}^{(t)}),  
~~c^{(t)}=-\text{SE}^{(t)}.
\vspace{-0.1cm}
\end{equation}  
Following the strategy parameterized by $\mu$, the long-term cumulative reward and cost starting from $t_{0}$ are
\vspace{-0.15cm}
\begin{equation}
    R(\mu)\!=\!\mathbb{E}_{\mu}\!\left[\!\sum_{t=t_{0}}^{T}\gamma^{t\!-\!t_{0}}r^{(t)}\!\right], ~
    C(\mu)\!=\!\mathbb{E}_{\mu}\!\left[\!\sum_{t=t_{0}}^{T}\gamma^{t\!-\!t_{0}}c^{(t)}\!\right],
    \vspace{-0.15cm}
\end{equation}
where $\gamma$ denotes the discounting factor. The objective of BS in this CMDP is to find
\vspace{-0.3cm}
\begin{align}
    \mu^{*}&=\underset{ \mu }{\arg\max} ~R(\mu) \label{Problem3}   \\
    \vspace{-0.2cm}
    \text{s.t.} &~ C(\mu)\leqslant \Gamma_{\text{c}}=\sum_{t=t_{0}}^{T}-\gamma^{t-t_{0}}\eta_{\text{c}},  \tag{\ref{Problem3}{a}} \label{Problem3_a}
\vspace{-0.6cm}
\end{align}
where $\eta_{\text{c}}=\frac{U\log_{2}(1+\tau)}{T_{\text{s}}\Delta f}$ is the threshold for SE in each subframe, corresponding to the constraint in (\ref{Problem0_a}). 

\subsubsection{State transition}

In the $(t\!+\!1)$-th subframe, the state $\bm s^{(t+1)}\!\!=\!\!\{\bm S_{\text{H}}^{(t+1)},\! \bm S_{\text{P}}^{(t+1)}\}$ is evolved as $\bm S_{\text{H}}^{(t+1)}=\bm S_{\text{H}}^{(t)}$, and $\bm S_{\text{P}}^{(t+1)}=\left[ \bm P^{(t-2)}, \bm P^{(t-1)}, \bm P^{(t)} \right]$. The channel-related state remains unaltered, while the position-related state can be refreshed with the latest estimation, $\bm P^{(t)}$, through the BS-end echo processing at $t$.
The evolution of the position spectrum $\bm P^{(t)}$ across subframes is contingent upon the real positions of the users/scatterers, and the target, which are parameterized by [$\theta^{(t)}_{\text{u},\text{p}}$, $\tau^{(t)}_{\text{u},\text{p}}$] and [$\theta^{(t)}_{\text{s}}$, $d^{(t)}_{\text{s}}$] respectively. These parameters vary according to respective velocities $\bm v_{\text{u},\text{p}}$ and $\bm v_{\text{s}}$, where $\vert \bm v_{\text{u},\text{p}}\vert,\vert\bm v_{\text{s}}\vert\!\sim \!\mathcal{U}[v_{\text{min}},v_{\text{max}}]$ with the angle randomly selected from 0 and $\pi$.  The specific evolution process characterized by kinematic equations can be referred to \cite{dynamic_MP}.


\subsection{Learning Algorithm}

The number of feasible actions grows exponentially with the number of users, posing challenges for training. 
Commonly-used algorithms such as Deep-Q-Network (DQN) experience performance degradation when dealing with this high-dimensional action space \cite{DDPG}. The constraint on the cost further escalates the challenge of maximizing the reward.
To address these difficulties, we adopt the actor-critic based learning framework with a primal-dual updating, as well as a Wolpertinger-based action selection scheme.
The key components of the learning algorithm are as follows:

\subsubsection{Learning architecture}
To tackle this constrained optimization problem, the reward and cost critic networks, parameterized by $\phi_{\text{R}}$ and $\phi_{\text{C}}$, map from the joint state space and action space to reward and cost respectively, playing a role of action evaluation: $Q_{\text{R}}(\bm s^{(t)},\bm a^{(t)} \vert \phi_{\text{R}})$ and $Q_{\text{C}}(\bm s^{(t)},\bm a^{(t)} \vert \phi_{\text{C}})$. 
The actor network $\mu$, parameterized by $\phi_{\text{A}}$, maps from the state space to an action $\bm a\in \mathbb{R}^{N_{\text{A}}}$ as $\mu(\bm s^{(t)} \vert \phi_{\text{A}})=\bm a^{(t)}$.
To reduce fluctuations in target values during training and expedite convergence, we employ target networks $Q_{\text{R}}'$, $Q_{\text{C}}'$ and $\mu'$ parameterized by $\phi_{\text{R}}'$, $\phi_{\text{C}}'$ and $\phi_{\text{A}}'$ respectively.
To solve the considered CMDP in (15), we first adopt the Lagrangian relaxation procedure to transform the problem in (\ref{Problem3}) to an unconstrained one as
\vspace{-0.3cm}
\begin{equation}
    \underset{ \lambda \geqslant 0 }{\min}~ \underset{ \phi_{\text{A}} }{\max}~ \left [R(\phi_{\text{A}})-\lambda (C(\phi_{\text{A}})-\Gamma_{\text{c}}) \right ],
    \label{unconstrained}
    \vspace{-0.2cm}
\end{equation}
where the dual variable $\lambda$ is another parameter to be updated.

\begin{figure}[t]
  \vspace{-0.4cm}
  \setlength{\abovecaptionskip}{-0.2cm} 
  \setlength{\belowcaptionskip}{-1cm} 
  \centering 
  \includegraphics[width=0.90\linewidth]{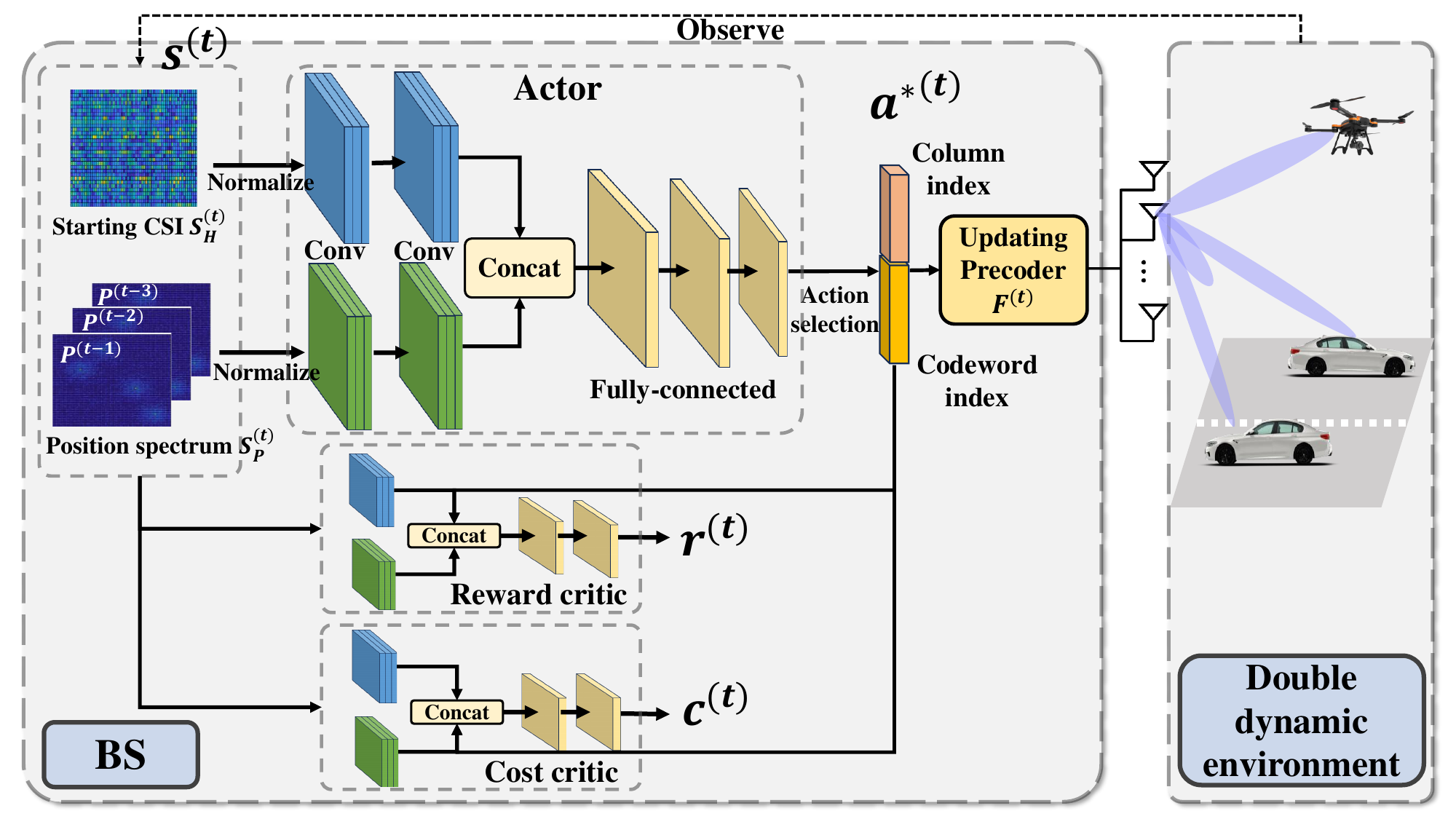}
  \caption{The proposed CDRL-based framework for ISAC precoding.}
  \label{fig:framework_conf}
  \vspace{-0.6cm}
\end{figure}

\subsubsection{Action selection}
In practice, the number of serving users may change. Thus, we resort to a dedicated action selection scheme as operated in the Wolpertinger architecture \cite{Wolpertinger} to facilitate a flexible training update. 
\begin{itemize}
    \item \textbf{Action exploration}: In order to prevent BS from getting stuck in local optima due to insufficient exploration, additional noise is added as to $\bm a^{(t)}$: $\tilde{\bm a}^{(t)}\!=\!\bm a^{(t)}\!+\mathcal{N}$, with $\mathcal{N}$ being the Ornstein-Uhlenbeck noise \cite{Wolpertinger}. 
    \item \textbf{Action quantization}: We first map $\widetilde{\bm a}^{(t)}$ into a set of binary sequences, $\hat{\mathcal{A}}$, followed by implementing $K$-nearest neighbors (KNN) algorithm to obtain $K$ feasible candidates closest to $\widetilde{\bm a}^{(t)}$ in Euclidean space.
    \item \textbf{Action refinement}: With the dual variable $\lambda$ fixed, the objective of BS is to maximize the unconstrained function in Eq.~(\ref{unconstrained}). Consequently, one action $\bm a^{*}$ is selected from the $K$ candidates and executed by BS at $t$, based on the current evaluation from the critic networks.
\end{itemize}
\begin{equation}
\setlength{\abovedisplayskip}{2pt}
\setlength{\belowdisplayskip}{2pt}
    {\bm a^{*}}^{(t)}\!=\!\mathop{\arg\max}\limits_{\bm a\in \hat{\mathcal{A}}}\! \left [ Q_{\text{R}}(\bm s^{(t)}\!, \bm a \vert \phi_{\text{R}})\!-\!\lambda (Q_{\text{C}}(\bm s^{(t)}\!, \bm a \vert \phi_{\text{C}}\!)\!-\!\Gamma_{\text{c}}) \!\right ].\label{equ:action selection}
    \vspace{-0.2cm}
\end{equation}
The transition tuple $(\bm s^{(t)}, \bm a^{(t)}, r^{(t)}, c^{(t)},\bm s^{(t+1)})$ is stored in a memory replay buffer $\mathcal{M}$ for network updating.

\subsubsection{Primal-Dual Network Update}
To facilitate the model training, we adopt a primal dual-deep deterministic policy gradient (PD-DDPG) procedure, where the policy parameterized by $\phi_{\text{A}}$ and the dual variable $\lambda$ are updated alternately for objective maximization while the critics $\phi_{\text{R}}$ and $\phi_{\text{C}}$ are updated for more precise policy evaluation.
To be specific, a batch of $N_{\text{b}}$ transition tuples are randomly sampled from $\mathcal{M}$, and the target reward and target cost is calculated as
\vspace{-0.2cm}
\begin{equation}
    \begin{aligned}
        y_{i}&=r_{i}+\gamma Q'_{\text{R}}(\bm s_{i+1}, \mu'(\bm s_{i+1} \vert \phi_{\text{A}}' )\vert \phi_{\text{R}}'),  \\
    z_{i}&=c_{i}+\gamma Q'_{\text{C}}(\bm s_{i+1}, \mu'(\bm s_{i+1} \vert \phi_{\text{A}}' )\vert \phi_{\text{C}}'),
    \label{equ:value_target}
    \end{aligned}
    \vspace{-0.2cm}
\end{equation}
where $i=1, \cdots, N_{\text{b}}$. Then the reward and cost critic networks are updated by minimizing the mean square error
\vspace{-0.2cm}
\begin{equation}
    \begin{aligned}
        L_{\text{R}}&=\frac{1}{N}\sum_{i}(y_{i}-Q_{\text{R}}(\bm s_{i}, \bm a_{i}\vert \phi_{\text{R}}))^{2},  \\
    L_{\text{C}}&=\frac{1}{N}\sum_{i}(z_{i}-Q_{\text{C}}(\bm s_{i}, \bm a_{i}\vert \phi_{\text{C}}))^{2},
    \label{equ:update_critic}
    \end{aligned}
    \vspace{-0.2cm}
\end{equation}
with the learning rate $\kappa_{\text{R}}$ and $\kappa_{\text{C}}$ respectively, followed by the actor updating through the sampled policy gradient ascend as 
\vspace{-0.2cm}
\begin{equation}
    \phi_{\text{A}}^{(t+1)}=\phi_{\text{A}}^{(t)}+\kappa_{\text{A}}\nabla_{\phi_{\text{A}}}\left [ R(\phi_{\text{A}})-\lambda (C(\phi_{\text{A}})-\Gamma_{\text{c}})\right ],
    \label{equ:update_actor}
\end{equation}
with the gradient being $\!\frac{1}{N}\!\sum_{i}\!\nabla_{\phi_{\text{A}}}\![Q_{\text{R}}(\bm s, \mu(\bm s \vert \phi_{\text{A}})\vert \phi_{\text{R}})\!-\!\lambda Q_{\text{C}}\!(\bm s, \mu(\bm s \vert \phi_{\text{A}})\vert \phi_{\text{C}}) ]\vert_{\bm s=\bm s_{i}}$ and the learning rate as $\kappa_{\text{A}}$.
Next the dual variable is updated by gradient descend as 
\vspace{-0.1cm}
\begin{equation}
    \lambda^{(t+1)}=\max( \lambda^{(t)}+ \kappa_{\lambda}\nabla_{\lambda} ,0),
\label{equ:update_dual}
\end{equation}
with $\nabla_{\lambda}=\frac{1}{N}\sum_{i}\left [Q_{\text{C}}(\bm s_{i}, \mu(\bm s_{i}\vert \phi_{\text{A}}))-\Gamma_{\text{c}}\right ]$ and the step size $\kappa_{\lambda}$.
Finally, we perform the soft updating of the target networks as \cite{DDPG}.
The pseudo code of the training stage of the proposed scheme is illustrated in Alg.~1.
After completing the offline training, the proposed scheme applies the trained actor to perform online updates of $\bm F$. The framework of the proposed precoding scheme is demonstrated in Fig.~\ref{fig:framework_conf}

In the learning architecture of the actor and critics, convolutional neural networks (CNN) are employed for processing $\bm S_{\text{H}}$ and $\bm S_{\text{P}}$ respectively, then the features are concatenated followed by fully-connected networks (FCN). Denote $L_{\text{c}}$ as the maximal number of convolutional layers, $N_{\text{f}}$ as the maximal number of input and output feature maps, and $N_{\text{k}}$ as the maximal side length of the filters in CNN.
$N_{\text{L}}$ and $N_{\text{w}}$ as the maximal number of neurons in the hidden layers and the number of hidden layers in FCN. 
Then the complexity in the forward pass of the actor network in online deployment is $\mathcal{O}(L_{\text{c}}(UG_{\text{t}}N_{\text{d}}+N_{\text{x}}N_{\text{y}})N_{\text{k}}^{2}N_{\text{f}}^{2} +N_{\text{w}}^{2}N_{\text{L}})$.


\vspace{-0.2cm}
\begin{algorithm}[h]
  \caption{CDRL-aided ISAC Precoding}
  \label{alg:ddpg}
  \begin{algorithmic}[1]
    \State Initialize actor network $\mu(\bm s \vert \phi_{\text{A}})$, reward critic network $Q_{\text{R}}(\bm s, \bm a \vert \phi_{\text{R}})$ and cost critic network $Q_{\text{C}}(\bm s, \bm a \vert \phi_{\text{C}})$ with random weights $\phi_{\text{A}}$, $\phi_{\text{R}}$ and $\phi_{\text{C}}$;
    \State Initialize target networks $\phi_{\text{A}}'\!=\!\phi_{\text{A}}$, $\phi_{\text{R}}'\!=\!\phi_{\text{R}}$ and $\phi_{\text{C}}'\!=\!\phi_{\text{C}}$;
    \State Initialize replay buffer $\mathcal{M}=\varnothing$, dual variation $\lambda=0$, mini-batch size $N_{\text{b}}=64$ and discount factor $\gamma=0.6$;
    \State Initialize a random Ornstein-Uhlenbeck process $\mathcal{N}$;
    \State Set initial state $\bm s^{(0)}$;
    \For{$t=0$ to $T$}
      \State Receive an action $\bm a^{(t)}=\mu(\bm s^{(t)}\vert \phi_{\text{A}})$ from $\mu(\bm s^{(t)}\vert \phi_{\text{A}})$;
      \State Add exploration noise $\mathcal{N}$, and select a binary sequence $~~~~~~{\bm a^{*}}^{(t)}$ as (17);
      \State Update $\bm F^{(t)}$ and observe $r^{(t)}$, $c^{(t)}$ and $\bm s^{(t+1)}$;
      \State Store the transition $(\bm s^{(t)}, \bm a^{(t)},r^{(t)}, c^{(t)},\bm s^{(t+1)})$ in $\mathcal{M}$;
      \State Sample a random batch of $N_{\text{b}}$ transitions from $\mathcal{M}$;
      \State Compute the target values $y_{i}$ and $z_{i}$ as (\ref{equ:value_target});
      \State Update the critic networks according to (\ref{equ:update_critic});
      \State Update the actor network according to (\ref{equ:update_actor});
      \State Update dual variable as (\ref{equ:update_dual});
      \State Update the target network parameters;
    \EndFor
  \end{algorithmic}
\end{algorithm}

\section{Experiments}
\vspace{-0.2cm}
In this section, simulation results are presented to evaluate the performance of the proposed scheme.
The adopted dataset incorporates both time-varying channel data and moving target information to reflect practical double dynamics \cite{dataset_M3SC}.
Unless otherwise specified, the system parameters are set as: $N_{\text{t}}=N_{\text{r}}=32$, $M=32$, $L_{\text{cp}}=8$, $L=32$, $T=100$, $U^{\text{max}}=16$, $N_{\text{B}}=32$, $G_{\text{t}}=64$. The central frequency point is set as $f_{\text{c}}=28$ GHz with interval $\Delta f=30$ kHz. $T_{\text{s}}=\frac{1}{\Delta f}=33.3~\mu$s. $g_{\text{rp}}(\cdot)$ is set as the raised-cosine filter with roll-off factor as 0.4. The channel-related parameters are set as $N_{\text{d}}=8$, $P_{\text{u}}=4$, $\sigma_{\beta}=1$. 
$\alpha_{\text{m}}$ has a random phase uniformly distributed in $[0,2\pi)$, and its amplitude obeys that $\vert \alpha_{\text{m}} \vert (\text{dB})=5\log_{10}(\sigma_{\text{RCS}})-10\log_{10}(f_{\text{m}})-20\log_{10}(d)+110$ according to \cite{radar_gain_model}, where $\sigma_{\text{RCS}}=10$ is the RCS of the target.
The noise variances are set as $\sigma_{\text{c}}^{2}=\sigma_{\text{z}}^{2}=-10$ dBm. SNR is defined as $\frac{P_{\text{t}}}{\sigma_{\text{c}}^{2}}$.
In the experiment environment, the sensing range is $[-\pi, \pi)\times [ 0~ \text{m}, 50~\text{m})$. The velocity (m/s) of the target, the users and the related scatters obey a uniform distribution of $\mathcal{U}[10,30]$. 
Main hyper-parameters for model training in experiments are set as: the size of the replay buffer $\mathcal{M}$ is $10^5$, the batch size is $N_{\text{b}}=64$, the learning rates of the actor, reward critic and cost critic are $(\kappa_{\text{A}},\kappa_{\text{R}},\kappa_{\text{C}})=(0.05,0.05,0.1)$. The discounting factor is $\gamma=0.6$, the parameters of the O-U noise for exploration is $(\mu_{\text{OU}},\theta_{\text{OU}},\sigma_{\text{OU}})=(0,0.5,0.3)$. We adopt Adam as the optimizer.
CDRL-PC and CDRL-PO denote the proposed algorithm with different state inputs respectively (positions plus initial CSI, positions-only). The perfect prior means that the instantaneous CSI and precise positions are available at BS.
Main benchmarks include the optimization-based method in Sec. III, greedy selection in \cite{shijian_MI} and random selection.
The proposed algorithm is developed using PyTorch and on a Window™-based machine running an NVIDIA™ GeForce RTX4060 GPU.

\begin{figure}[t]
  \vspace{-0.1cm}
  \setlength{\abovecaptionskip}{0cm} 
  \setlength{\belowcaptionskip}{-1cm} 
  \centering
   \subfigure[Cumulative reward.\label{fig:reward}]
    {\includegraphics[width=0.46\linewidth]{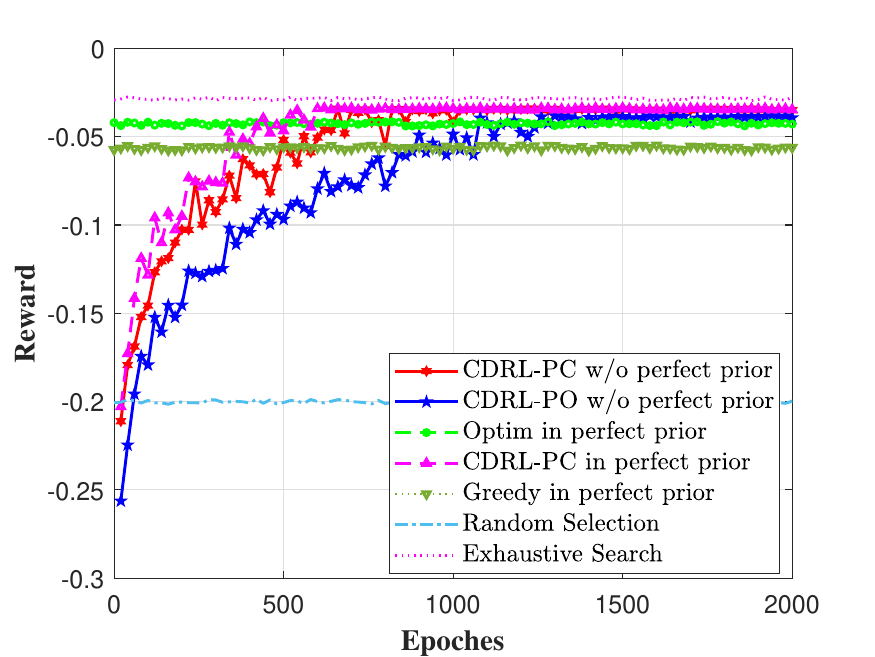}}
    \hspace{0.1cm}
  \subfigure[Cumulative cost.\label{fig:rate}]
    {\includegraphics[width=0.46\linewidth]{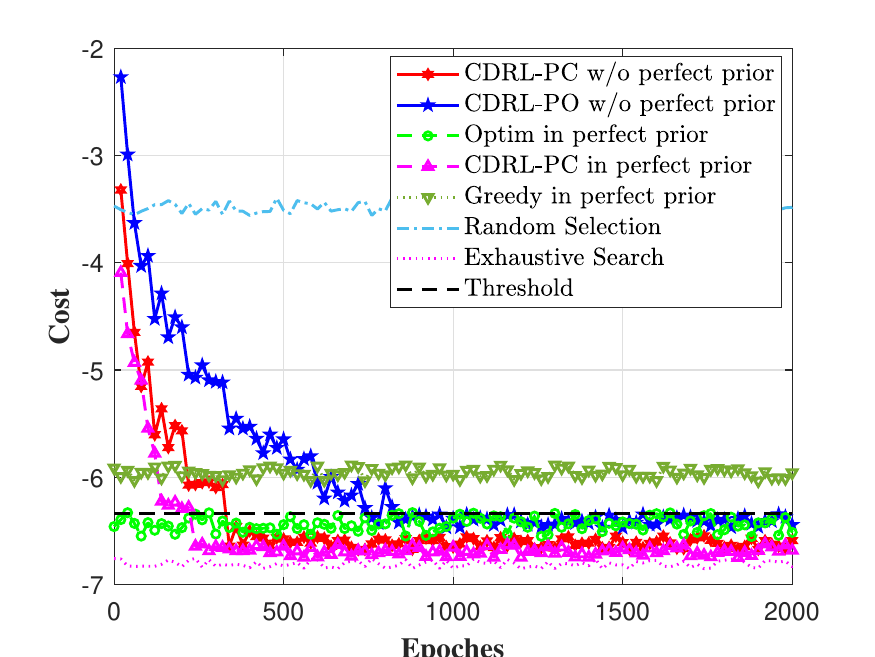}}
    \vspace{-0.1cm}
  \caption{Cumulative reward and cost versus epochs.}
  \vspace{-0.6cm}
  \label{fig:reward}
\end{figure}


Firstly, we verify the agent’s achievable cumulative reward and cost within one frame in the training process of the proposed scheme in Fig.~3. We set the SINR threshold in (\ref{Problem0_a}) as $\tau=3$ dB and $U=4$.
As the number of episodes increases, the cumulative reward of the proposed algorithm gradually rises, and BS progressively learns to satisfy the SE constraint, resulting in the cumulative cost falling below the threshold. After convergence, the cumulative reward can closely approach the scenario with perfect prior and optimal exhaustive search, surpassing other benchmarks while ensuring the SE requirement. This reflects the proposed CDRL-aided method is less dependent on perfect prior.


We then verify the sensing performance of the proposed scheme in Fig.~4(a). Despite lacking perfect prior information, the proposed scheme achieves a lower averaged CRLB than existing benchmarks, exhibiting a performance gain of approximately 2~dB compared to optimization-based method. Even when relying solely on historical position estimates in CDRL-PO version, the proposed scheme maintains high accuracy.
Furthermore, we analyze the impact of user quantity on SE at an SNR of 0 dB in Fig.~4(b). The average SE initially improves with an increase in users but declines beyond a certain threshold, as the increase of $U$ exacerbates IUI. 
Leveraging Wolpertinger-based action selection, the proposed algorithm demonstrates superior adaptability to varying user numbers, preventing the SE crash observed in non-Wolpertinger DQN-based algorithm under high user loads.


\begin{figure*}[t]
  \vspace{-0.64cm}
  \setlength{\abovecaptionskip}{0cm} 
  \setlength{\belowcaptionskip}{-1cm} 
  \centering
   \subfigure[Averaged CRLB versus SNR.\label{fig:crb}]
    {\includegraphics[width=0.30\linewidth]{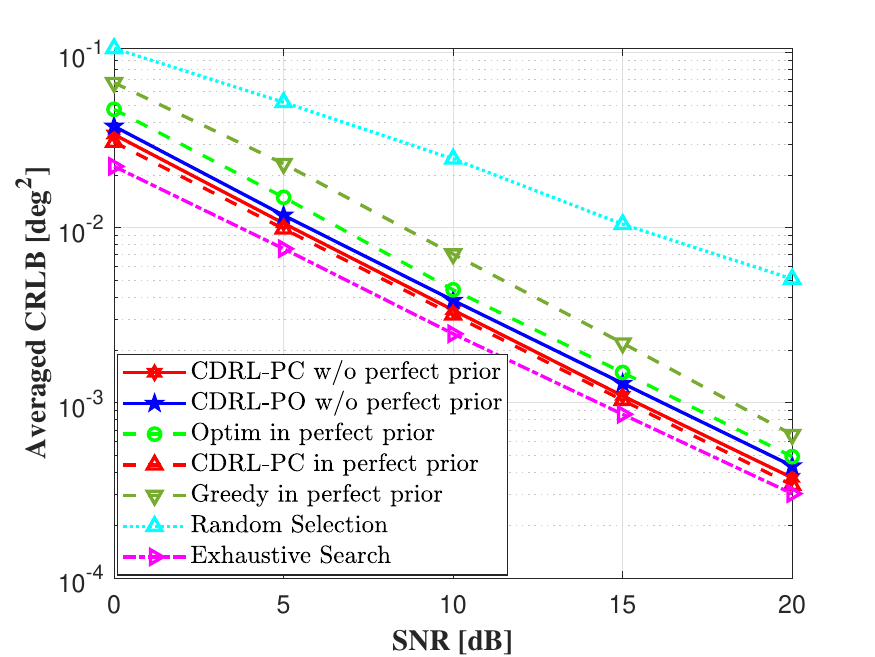}}
    \hspace{0.1cm}
  \subfigure[Averaged SE versus number of users.\label{fig:rate}]
    {\includegraphics[width=0.30\linewidth]{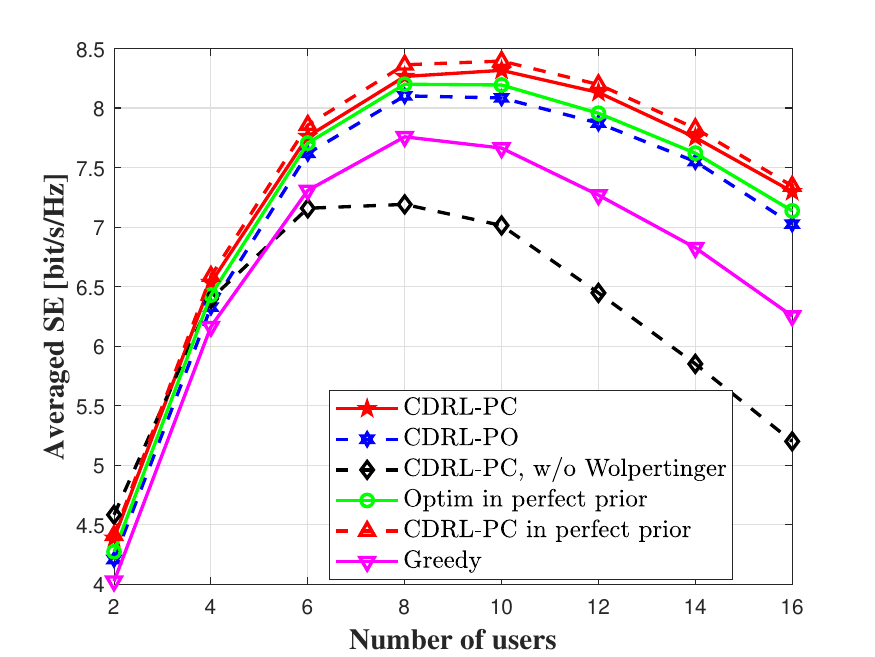}}
    \hspace{0.1cm}
    \subfigure[ISAC performance region.\label{fig:bound}]
    {\includegraphics[width=0.30\linewidth]{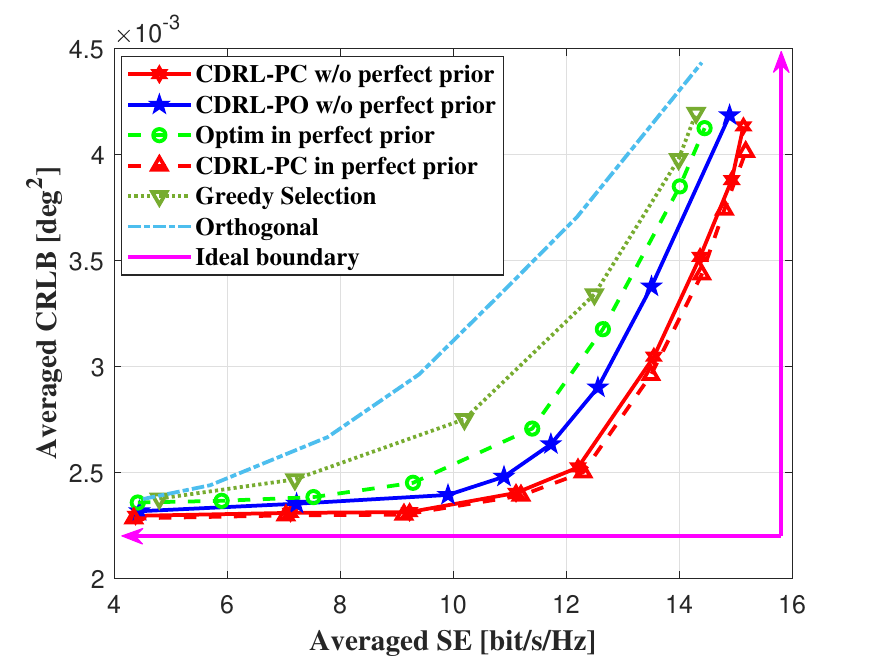}}
    \vspace{-0.1cm}
  \caption{ISAC performance comparison among various schemes.}
  \vspace{-0.65cm}
  \label{fig:ISAC}
\end{figure*}

As a step further, we describe the achievable ISAC performance boundary in Fig.~4(c).
The proposed scheme achieves a significant gain compared to the existing benchmarks in terms of tradeoff, approaching the ideal boundary. Compared to that under the completely orthogonal operations where the communication and sensing are separate, the achieved bound is considerably enhanced, reflecting the cooperation gain of ISAC, as analyzed in \cite{JCR_ISIT}.
The proposed approach reduces complexity compared to other methods, as shown in Table.~1.

\vspace{-0.5cm}
\begin{table}[h]
    \centering
    \caption{Complexity comparison}
    \vspace{-0.2cm}
    \begin{tabular}{|p{1.8cm}|p{4.9cm}|}
    \hline
       Method  &  Complexity \\
       \hline
       \hline
       Proposed  &  $\mathcal{O}(L_{\text{c}}(UG_{\text{t}}N_{\text{d}}\!+\!N_{\text{x}}N_{\text{y}})\!N_{\text{k}}^{2}N_{\text{f}}^{2} \!+\!N_{\text{w}}^{2}N_{\text{L}})$       \\
       \hline
       Optimization  &  $\mathcal{O}(N_{\text{t}}^{3.5}\log(1/\epsilon))$     \\
       \hline
       Greedy &  $\mathcal{O}(U^{4}N_{\text{t}}M)$      \\
       \hline
       Exhaustive  &  $\mathcal{O}((N_{\text{B}})^{\text{U}})$      \\
       \hline
    \end{tabular}
    \label{tab:parameter}
    \vspace{-0.3cm}
\end{table}

\vspace{-0.2cm}

\section{Conclusions}

\vspace{-0.2cm}


In this paper, we proposed a CDRL approach to tackle the ISAC precoding problem in scenarios where both communication and sensing exhibit time variation. By utilizing various state observations such as the historical positions and initial CSI, the proposed scheme effectively adapts to the double dynamics. Moreover, the enhancement of our training algorithm allows us to effectively address constrained problems and seamlessly adapt to varying user numbers. The proposed scheme showcases enhancements in performance boundaries within double dynamics, achieving a minimum of 2 dB performance gain in sensing while ensuring SE for communications. It also diminishes reliance on perfect prior information and computational complexity, making it a promising solution for supporting vehicular networks.






\ifCLASSOPTIONcaptionsoff
  \newpage
\fi

\vspace{-0.2cm}
\bibliographystyle{IEEEtran}
\bibliography{IEEEabrv,myrefs}

\begin{thebibliography}{10}
\providecommand{\url}[1]{#1}
\csname url@samestyle\endcsname
\providecommand{\newblock}{\relax}
\providecommand{\bibinfo}[2]{#2}
\providecommand{\BIBentrySTDinterwordspacing}{\spaceskip=0pt\relax}
\providecommand{\BIBentryALTinterwordstretchfactor}{4}
\providecommand{\BIBentryALTinterwordspacing}{\spaceskip=\fontdimen2\font plus
\BIBentryALTinterwordstretchfactor\fontdimen3\font minus \fontdimen4\font\relax}
\providecommand{\BIBforeignlanguage}[2]{{%
\expandafter\ifx\csname l@#1\endcsname\relax
\typeout{** WARNING: IEEEtran.bst: No hyphenation pattern has been}%
\typeout{** loaded for the language `#1'. Using the pattern for}%
\typeout{** the default language instead.}%
\else
\language=\csname l@#1\endcsname
\fi
#2}}
\providecommand{\BIBdecl}{\relax}
\BIBdecl

\bibitem{overview_JRC_for_AV}
D.~Ma, N.~Shlezinger, T.~Huang, Y.~Liu, and Y.~C. Eldar, ``Joint {Radar-Communication} strategies for autonomous vehicles: {Combining} two key automotive technologies,'' \emph{{IEEE} Signal Process. Mag.}, vol.~37, no.~4, pp. 85--97, Jul. 2020.

\bibitem{cheng2022integrated}
X.~Cheng, D.~Duan, S.~Gao, and L.~Yang, ``Integrated sensing and communications ({ISAC}) for vehicular communication networks ({VCN}),'' \emph{{IEEE} Internet Things J.}, vol.~9, no.~23, pp. 23\,441--23\,451, Dec. 2022.

\bibitem{Joint_opt_ISAC}
X.~Liu \emph{et~al.}, ``Joint transmit beamforming for multiuser {MIMO} communications and {MIMO} radar,'' \emph{{IEEE} Trans. Signal Process.}, vol.~68, pp. 3929--3944, Jun. 2020.

\bibitem{radar_integrated_V2V_Fan}
Y.~Fan, S.~Gao, D.~Duan, X.~Cheng, and L.~Yang, ``Radar integrated {MIMO} communications for multi-hop {V2V} networking,'' \emph{{IEEE} Wireless Commun. Lett.}, vol.~12, no.~2, pp. 307--311, Feb. 2023.

\bibitem{ISAC_CRB_opt}
F.~Liu, Y.-F. Liu, A.~Li, C.~Masouros, and Y.~C. Eldar, ``{Cramér-Rao} bound optimization for joint radar-communication beamforming,'' \emph{{IEEE} Trans. Signal Process.}, vol.~70, pp. 240--253, Dec. 2022.

\bibitem{overview_Som}
X.~Cheng \emph{et~al.}, ``Intelligent multi-modal sensing-communication integration: Synesthesia of machines,'' \emph{{IEEE} Commun. Surveys Tuts.}, vol.~26, no.~1, pp. 258--301, 1st Quart. 2023.

\bibitem{MMFF_Som}
H.~Zhang, S.~Gao, X.~Cheng, and L.~Yang, ``Integrated sensing and communications towards proactive beamforming in {mmWave} {V2I} via multi-modal feature fusion ({MMFF}),'' \emph{{IEEE} Trans. Wireless Commun.}, 2024, early access.

\bibitem{dynamic_com_1}
Z.~Gao \emph{et~al.}, ``Integrated sensing and communication with {mmWave} massive {MIMO}: {A} compressed sampling perspective,'' \emph{{IEEE} Trans. Wireless Commun.}, vol.~22, no.~3, pp. 1745--1762, Mar. 2023.

\bibitem{dynamic_com_2}
Y.~Li, X.~Yuan, Y.~Hu, J.~Yang, and A.~Schmeink, ``Optimal {UAV} trajectory design for moving users in {ISAC} networks,'' \emph{{IEEE} Trans. Intell. Transp. Syst.}, vol.~24, no.~12, pp. 15\,113--15\,130, Dec. 2023.

\bibitem{dynamic_sensing_2}
T.~P. Zieliński \emph{et~al.}, ``Wireless {OTFS}-based {ISAC} for moving vehicle detection,'' \emph{{IEEE} Sensors J.}, vol.~24, no.~5, pp. 6573--6583, Mar. 2024.

\bibitem{dynamic_MP}
W.~Yuan \emph{et~al.}, ``Bayesian predictive beamforming for vehicular networks: {A} low-overhead joint radar-communication approach,'' \emph{IEEE Trans. Wireless Commun.}, vol.~20, no.~3, pp. 1442--1456, Mar. 2021.

\bibitem{dynamic_PF}
Z.~Ying, Y.~Cui, J.~Mu, and X.~Jing, ``Particle filter based predictive beamforming for integrated vehicle sensing and communication,'' in \emph{IEEE 94th Vehicular Technology Conference (VTC2021-Fall)}, 2021, pp. 1--5.

\bibitem{RL_radar_ISAC}
W.~Zhai, X.~Wang, M.~S. Greco, and F.~Gini, ``Reinforcement learning based integrated sensing and communication for automotive {MIMO} radar,'' in \emph{2023 IEEE Radar Conference (RadarConf23)}, 2023, pp. 1--6.

\bibitem{RL_time_divide}
L.~Xu, R.~Zheng, and S.~Sun, ``A {DRL} approach for integrated automotive radar sensing and communication,'' in \emph{IEEE 12th Sensor Array and Multichannel Signal Processing Workshop (SAM)}, 2022, pp. 316--320.

\bibitem{SAC}
T.~Haarnoja \emph{et~al.}, ``Soft actor-critic: {Off}-policy maximum entropy deep reinforcement learning with a stochastic actor,'' \textit{arXiv preprint arrXiv:1801.01290}, 2018.

\bibitem{DDPG}
T.~P. Lillicrap \emph{et~al.}, ``Continuous control with deep reinforcement learning,'' \textit{arXiv preprint arrXiv:1509.02971}, 2019.

\bibitem{DSDS_TWC}
S.~Gao, X.~Cheng, and L.~Yang, ``Estimating {Doubly-Selective} channels for hybrid {mmWave} massive {MIMO} systems: {A} doubly-sparse approach,'' \emph{{IEEE} Trans. Wireless Commun.}, vol.~19, no.~9, pp. 5703--5715, Sept. 2020.

\bibitem{fan2021wideband}
Y.~Fan, S.~Gao, X.~Cheng, L.~Yang, and N.~Wang, ``Wideband generalized beamspace modulation for {mmWave} massive {MIMO} over doubly-selective channels,'' \emph{IEEE Trans. Veh. Technol.}, vol.~70, no.~7, pp. 6869--6880, Jul. 2021.

\bibitem{location_aware_overview}
M.~Koivisto \emph{et~al.}, ``High-efficiency device positioning and location-aware communications in dense {5G} networks,'' \emph{{IEEE} Commun. Mag.}, vol.~55, no.~8, pp. 188--195, Aug. 2017.

\bibitem{Wolpertinger}
G.~Dulac-Arnold \emph{et~al.}, ``Deep reinforcement learning in large discrete action spaces,'' 2016.

\bibitem{dataset_M3SC}
X.~Cheng \emph{et~al.}, ``{M3SC}: A generic dataset for mixed multi-modal sensing and communication integration,'' \emph{China Commun.}, vol.~20, no.~11, pp. 13--29, 2023.

\bibitem{radar_gain_model}
Z.~Chai \emph{et~al.}, ``Empirical path loss channel modeling at 28 {GHz} for isac system,'' in \emph{2023 International Conference on Wireless Communications and Signal Processing (WCSP)}, 2023, pp. 366--370.

\bibitem{shijian_MI}
S.~Gao, X.~Cheng, and L.~Yang, ``Mutual information maximizing wideband multi-user {(wMU)} {mmWave} massive {MIMO},'' \emph{{IEEE} Trans. Commun.}, vol.~69, no.~5, pp. 3067--3078, May 2021.

\bibitem{JCR_ISIT}
M.~Kobayashi, H.~Hamad, G.~Kramer, and G.~Caire, ``Joint state sensing and communication over memoryless multiple access channels,'' in \emph{2019 IEEE International Symposium on Information Theory (ISIT)}, 2019, pp. 270--274.

\end{thebibliography}

\end{document}